\documentclass[12pt,journal,onecolumn]{IEEEtran}

\usepackage{graphicx}
\usepackage{epstopdf}
\usepackage[cmex10]{amsmath}
\usepackage{amsfonts}
\usepackage[ruled,linesnumbered]{algorithm2e}
\usepackage{algorithmic}
\usepackage{array}
\usepackage{eqparbox}
\usepackage{epsfig}
\usepackage{color}
\usepackage{graphicx}
\usepackage{times}
\usepackage{float}
\usepackage{stfloats}
\usepackage{graphicx}
\usepackage{subfigure}
\usepackage{type1cm}
\usepackage{url}
\usepackage{cite}
\usepackage{enumerate}
\usepackage{lineno}
\usepackage{amsthm}
\usepackage{amssymb}
\usepackage{bbm}

\theoremstyle{definition}

\newcommand{\argmax}{\operatornamewithlimits{arg\,max}}

\begin{document}
%
\title{On Maximizing Energy and Data Delivery in Dense Wireless Local Area Networks} 
\author{Kwan-Wu Chin

\thanks{University of Wollongong, NSW, Australia. Email: kwanwu@uow.edu.au}}

\maketitle

\begin{abstract}
Devices can now be powered wirelessly by Access Points (APs). 
However, an AP cannot transmit frequently to charge devices as it may starve other nearby APs operating on the same channel.  Consequently, there is a need to schedule the transmissions of APs to ensure their data queues remain short whilst charging energy-harvesting devices.
We present a finite-horizon Markov Decision Process (MDP) to capture the queue states at APs and also channel conditions to nodes.  We then use the MDP to investigate the following transmission policies: max weight, max queue, best channel state and random.
Our results show that the max weight policy has the best performance in terms of queue length and  delivered energy.

\end{abstract}


%
\IEEEpeerreviewmaketitle

\section{\label{INTRO}Introduction}
Wireless Local Area Networks (WLANs) are now ubiquitous with Access Points (APs) densely deployed to ensure high network capacity \cite{Dense15}. In fact, small, dense cells will constitute future 5G networks \cite{Dense2014}.  In addition to providing superior capacity to nodes, densely deployed APs or cells are envisaged to help proliferate low-power devices that form the Internet of Things (IoTs).  This is because low-power devices such as sensor nodes are able to use transmissions from APs as a power source; examples of which, with a temperature sensor or camera, have been demonstrated in \cite{WPT1}.
%
%

\begin{figure}[htbp]
	\centering
	\includegraphics[scale=0.6]{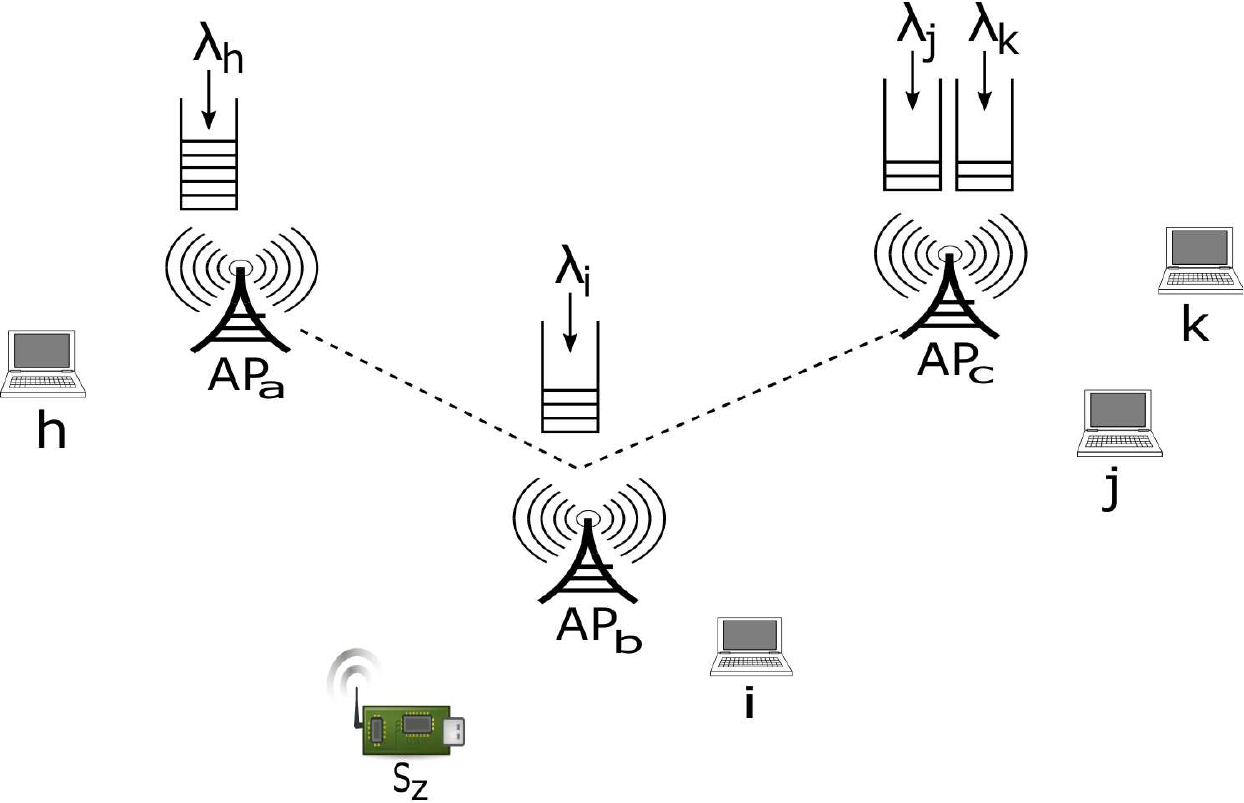}
	\caption{An example dense WLAN with an RF-charging sensor node $S_z$. Dotted lines indicate interference between APs. }
	\label{FIG1}
\end{figure}

The feasibility of RF-energy harvesting coupled with the use of dense cells have spurred the development of joint charging and data transmission schemes.  These works aim to supply energy to devices in order to maximize their throughput.  Examples include ``harvest-then-transmit'' approaches, where an AP first charges devices wirelessly.  These devices then set their transmission rate to the AP based on the amount of harvested energy.  In \cite{TPUTWPN}, the authors determine the energy harvesting time that maximizes the minimum transmission rate.
The same problem is then extended to cases where the AP has Multiple-Input Multiple-Output (MIMO) capability \cite{WPCNMIMO2}; for other extensions, please see \cite{RF1} and references therein.
In \cite{WPTQ1} the authors propose to adapt an AP's energy beamforming vector, transmission time and power of devices to ensure the queue at these devices remain stable. 
We emphasize that these works assume one AP and do not consider {\em multiple} interfering APs scenarios; e.g., in Figure \ref{FIG1}, all APs are on the same channel, meaning AP$_a$ and AP$_b$ cannot transmit simultaneously.  Hence, link scheduling is critical.  Also, these works do not consider queue dynamics at APs.  This is important as APs are responsible for delivering data.
In a different example, the authors of \cite{RFMACSK} design a signaling protocol that pairs sinks and sensor nodes, either for charging or data collection; both of which are carried out simultaneously.  
They, however, did not consider data delivery from APs.
In \cite{WPT1}, the authors prototype sensor nodes that harvest energy from APs.  They observe that APs traffic load is too low to charge sensor nodes.  A solution is to have APs transmit {\em power} or dummy User Datagram Protocol (UDP) packets when their queue occupancy is low.
The authors of \cite{WPTPROJ1} jointly optimize routing and link schedule over a multi-hop wireless network to ensure flow demands and energy requirement of energy harvesting sensor nodes are met.  
In the foregone works, all authors did not consider scheduling APs transmissions to minimize queue lengths or take advantage of favorable channel conditions.

Different from existing works, we consider the following system and problem.  Referring to Figure \ref{FIG1}, each AP has random exogenous data packet arrivals.  There is one sensor node $S_z$ that harvests energy whenever AP$_b$ transmits.  Hence, its energy level is directly linked to AP$_b$'s transmission duration, path loss and energy conversion efficiency $\alpha$.  In this example, we have the following transmission or independent sets $a_i=[AP_a,AP_b,AP_c]$: $a_1=[1,0,1]$ or $a_2=[0,1,0]$, where a `1' indicates a transmitting AP.  Assume time is slotted, where each slot lasts for 100ms.  One possible schedule is to activate $a_1$ and $a_2$ for 500ms (five slots) each.  This means for a received power $P_{bz}$, sensor node $S_z$ will harvest $\frac{1}{2}\alpha P_{bz}$ of energy.  Although this schedule activates APs equally, it may cause long queues; for example, if the active time or service rate of an AP, say $a$, is smaller than $\lambda_h$.  Conversely, a schedule that favors $a_1$ may starve $S_z$ of energy.  Also, when determining a schedule, we have to consider current and future channel conditions as well as data packet arrivals.
%

In summary, our problem is to determine a transmission schedule over $T$ time slots that optimizes the following reward or metrics: (i) queue lengths at APs, and (ii) energy harvested by sensor or energy harvesting nodes. 
Next, we present the finite-horizon Markov Decision Process (MDP) model used to study the following policies: (i) {\em max weight}, where APs select the queue with the biggest product value between queue length and channel condition, (ii) {\em max queue}, where APs select the station with the longest queue, (iii) {\em max channel state information (CSI)}, where APs select the station with the best channel, and (iv) {\em random}, where APs randomly return a queue to service.
Our results show that the max weight policy yields high arrival rates and delivered energy.
%

\section{Preliminaries}
\label{PROBD}
We denote a set of APs by $\mathcal{A}$; all of which operate on the same channel.  
A controller connects all APs and is aware of stations associated to each AP, and nearby sensor or energy-harvesting nodes.  
It also computes, using our approach in Section \ref{SOL}, and sets the transmission schedule for all APs every $T$ slots.  Each slot is indexed by $t$.  
%
APs serve two types of nodes: (i) {\em stations}, denoted as by the set $\mathcal{S}_d$, that transmit and receive data to/from their associated AP.  Stations do not rely on the APs for power.  Those associated to AP $a$ are placed in set $S_d(a)$, and 
(ii) the set of {\em sensor nodes}  $\mathcal{S}_e$. Those ``near'' AP $a$ are denoted as $\mathcal{S}_e(a)$, meaning when AP $a$ transmits, these sensor nodes are able to harvest energy because the received power is higher than a given received sensitivity; e.g., for the energy harvesters reported in \cite{RF1},  an RF input power in the range $-14$ to $-22$ dBm is sufficient to produce 1V DC output. 
Note, sensor nodes are able to harvest energy from one AP only.  This is reasonable given low receiver sensitivity that restricts the charging distance to no more than five meters \cite{WPT1}.
APs always have data packets; if not, they transmit power or UDP packets \cite{WPT1} to fill their allocated time slot.
%
%

APs may interfere due to the limited number of orthogonal channels in IEEE 802.11-based WLANs.  We use a conflict graph to represent the interference between APs.  Its vertices correspond to APs.
%
%
%
If two APs interfere, i.e., they hear each other's transmissions or an associated station hears transmissions from the other AP, then there is an edge in the conflict graph between the two APs.  
Note that the conflict graph is fixed for a given topology.  Also we do not consider uplink traffic.
With the conflict graph in hand, we can then derive the collection of transmission or independent sets; see Section \ref{EVAL} for an example heuristic.  Each of these sets then forms the columns of a  matrix $\textbf{A}$ with elements taking a value of zero or one.  For example, for the conflict graph of Figure \ref{FIG1}, namely AP$_a$-AP$_b$-AP$_c$, two possible transmission sets are $[1\; 0\; 1]^{T}$ and $[0\; 1\; 0]^T$; i.e., either AP$_a$ and AP$_c$ transmit together or AP$_b$ transmits by itself. Note, there can be up to $2^{|\mathcal{A}|}$ independent sets and finding the maximum independent set is an NP-hard problem.
In the sequel, $\tau(t)$ refers to the transmission set activated at time $t$.  Also, with a slight abuse of notation we will also treat $\textbf{A}$ as a set. 

Each AP has a First-In-First-Out (FIFO) queue for each associated station.  Formally, at AP $a$, the queue that stores packets headed to station $j$ at time $t$ is $Q_{aj}(t)$.  Let $A_{aj}(t)$ be the packet arrivals at time $t$ for station $j$.  Packets arrive at the start of each slot as per an i.i.d process with mean $\lambda_j$.   

The channel state evolves according to a finite-state Markov chain and is static for the duration of a time slot.  Moreover, the channel state to each station is independent and is known to the controller.  We write $\alpha_{az}(t)$ to represent the number of packets that can be received by station $z$ from AP $a$ at time $t$.  On the other hand, $\varepsilon_{az}(t)$ refers to the potential energy (in Joules) that can be harvested by sensor node $z$ at time $t$ for each packet transmitted by AP $a$.
%

In each time slot $t$, we assume APs select the {\em best} station according to some policy; see Section \ref{SOL}.  Let this station be $j^*$.
Assume AP $k\in\tau(t)$ is scheduled to transmit.  Its total number of transmitted packets is,
\begin{equation}
T_k(t) = MIN(Q_{kj^*}(t),\alpha_{kj^*}(t))
\end{equation}
The queue for each station $j$ at AP $k$ evolves as,
\begin{equation}
Q_{kj}(t+1) = Q_{kj}(t)+A_{kj}(t) - T_k(t)\mathbbm{1}_{j=j^*}
\end{equation}
In words, for each station $j$, we sum its current queue length plus any arrivals.  For station $j^*$, we subtract transmitted packets as the indicator function $\mathbbm{1}_x$ returns a value of one when $j$ equals $j^*$.  
%
%

Lastly, let $X_a(t)$ be the total queue length of AP $a$.  Formally, it evolves as per,
\begin{equation}
X_a(t+1) = \sum_{j\in S_d(a)} Q_{aj}(t+1)
\end{equation}

%

We ignore battery capacity because the energy delivered by an AP is a few orders of magnitude smaller than the battery capacity of sensor nodes; e.g., two AA batteries, as commonly used by sensor nodes, is capable of storing tens of kilo-joules of energy versus a recharging rate of micro-joules.

\section{A Markov Decision Process Model}
\label{MODEL}
Briefly, an MDP \cite{MDPBOOK} is specified by the tuple $(\mathbf{Y}, \mathbf{A}, P(.|.,.), r(.,.))$, where $\mathbf{Y}$ denotes the state space and $\mathbf{A}$ is the action space.  In state $\mathbf{y}\in\mathbf{Y}$, if we take action $a\in\mathbf{A}$, then we earn a reward $r(\mathbf{y},a)$.  After that, we move to a new state $\mathbf{y'}\in\mathbf{Y}$ with probability $P(\mathbf{y'}|\mathbf{y},a)$, where $P(\mathbf{y'}|\mathbf{y},a)\ge 0$ and $\sum_{\mathbf{y'}\in\mathbf{Y}} P(\mathbf{y'}|\mathbf{y},a)=1$.

We are now ready to instantiate a MDP for the problem at hand.
At time $t$, the queue length at all APs is recorded as $Q_t=(|Q_{aj}(t)|)_{a\in\mathcal{A}, j\in S_d}$.
The channel states, from an AP to sensor nodes and stations at time $t$, are stored in $C_t = (\alpha_{aj})_{a\in\mathcal{A}, j\in S_d}\cup (\alpha_{aj})_{a\in\mathcal{A}, j\in S_e}$.
The current state at time $t$ is denoted as $\mathbf{y}_t=(Q_t, C_t)$.  
Formally, we have a discrete-time Markov chain $\{\mathbf{y_t}\}_{t=0}^{t=\infty}$.
Lastly, we refer to $\mathbf{Y}=(Q, C)$ as the state space, where $Q$ and $C$ are the possible queue lengths and channel states, respectively.
Observe that the state space has a high dimensionality. For example, if the maximum queue length at each AP is $100$ packets and there are $10$ possible channel states, then in total we have $100^{|\mathcal{A}|}\times 10^{|\mathcal{S}_d|}$ states.
%

The action space corresponds to all transmission sets in matrix $\textbf{A}$.  
%
%
%
%
%
The transition probability to a new state $\mathbf{y_{t+1}}$, i.e., $P(\mathbf{y_{t+1}}\;|\;\mathbf{y_t}, \tau(t))$, given state $\mathbf{y_t}$ and action $\tau(t)$, is determined by both the Markov model that dictates the channel condition of each link and also the packet arrival process.

We now define the reward $r_t(\mathbf{y_t}, \tau(t))$. First, let the total number of packets transmitted by APs at time $t$ be, 
\begin{equation}
T(t) = \sum_{k\in\tau(t)} T_k(t)
\end{equation}
At time $t$, the total amount of energy delivered is
\begin{equation}
E(t) = \sum_{k\in\tau(t)}\sum_{j\in S_e(k)} \varepsilon_{kj}(t)T_k(t)
\end{equation}
The reward is then defined as,
\begin{equation}
r_t(\mathbf{y_t}, \tau(t)) = \gamma_dT(t)+ \gamma_eE(t)
\end{equation}
where $\gamma_d$ and $\gamma_e$ are weights to bias the reward.
Observe that the reward is tied closely to both the queue and channel state of the station chosen by each AP as well as the number of transmitting APs in $\tau(t)$.

%
%
Given $T$ time slots, our problem is to determine a {\em policy} $\pi$ that returns the best transmission set from $\mathbf{A}$ for each of the next $T$ slots such that the expected reward is maximized.  Formally, we have,
\begin{equation}
\label{MDPPROB}
\max_{\pi} \mathbb{E}\left[ \sum_{t=0}^{T} r_t(\mathbf{y_t},\pi(\mathbf{y_t}, \textbf{A})) \right]
\end{equation}
where the expectation $\mathbb{E}$ is taken with respect to random channel conditions and queue arrivals.
Note, in practice, the exact value of $T$ needs to be balanced against signaling overheads and computation time.  If $T$ is small, then frequent commands to APs may cause congestion.  On the other hand, if $T$ is big, then the computation time may be too long as the problem has high dimensionality. 
\section{Approximate DP and Policies}
\label{SOL}
We employ approximate DP given the high dimensionality of the problem at hand.  Specifically, we use forward dynamic programming, whereby for a given starting state $\mathbf{y}_0\in\mathbf{Y}$ and discount factor $\gamma$, the problem as formulated by (\ref{MDPPROB}) is equivalently to computing the following value function \cite{MDPBOOK},
\begin{align}
\label{MDPVAL}
V_t(\mathbf{y_t}) &= \max_{\tau(t)\in\mathbf{A}} \left[\vphantom{\sum_{\mathbf{y_{t+1}}\in\mathbf{Y}}} r_t(y_t, \tau(t))\; + \right.\nonumber &  \\
 & \left.\gamma\sum_{\mathbf{y_{t+1}}\in\mathbf{Y}}P(\mathbf{y_{t+1}}|\mathbf{y_{t}}, \tau(t))V_{t+1}(\mathbf{y_{t+1}})\right] &
\end{align}
A key problem is computing $V_{t}$ due the size of $\mathbf{Y}$.  To this end, we use approximate value iteration; see \cite{MDPBOOK}.  Specifically, we randomly generate a sample of $N$ outcomes from $\mathbf{Y}$.  Let the set $\Omega$ stores these $N$ outcomes, and $p(\omega)$ be the probability of outcome $\omega\in\Omega$.  We then rewrite Equ. (\ref{MDPVAL}) as, 
\begin{equation}
\label{MDPVALAPPROX}
\bar{V}_t(\mathbf{y_t}) = \max_{\tau(t)\in\mathbf{A}}\left[r_t(y_t, \tau(t))+\gamma\sum_{\omega\in\Omega}P(\omega)\bar{V}_{t+1}(\mathbf{\omega})\right]
\end{equation}
%

Implicitly in the calculation of $V_{t}$ (or $\bar{V}_t$) is that each AP has chosen the best station $j^*$. In this paper, we study the following policies.
The first is called {\em Max Weight}.  At each time $t$, AP $a$ selects station $j^*$ as follows, 
\begin{equation}
j^* = \argmax_{j\in S_d(a)}\; \alpha_{aj}(t)Q_{aj}(t)
\label{MAXW}
\end{equation}
That is, pick the station that has the longest queue and best channel.   
The second policy called {\em Max Queue} corresponds to APs picking only the longest queue, i.e., identify the station $j$ with the biggest $Q_{aj}(t)$ value.  The third policy, aka {\em Max CSI}, requires APs pick the station with biggest $\alpha_{aj}(t)$ value.  Lastly, the {\em Random} policy returns a data queue randomly.

%
%
\section{Evaluation}
\label{EVAL}
Using the formulated MDP, we now study the foregone policies as follows.  
We first study the following scenario.
There are $|\mathcal{A}|=10$ APs; each of which is randomly assigned up to five sensor nodes and five stations. 
In each simulation run, a new topology is generated; i.e., one with a different conflict graph, and APs have a new set of stations and sensor nodes.
The channel state at each time $t$ is driven by a Markov model. For sensor nodes, each state is the amount of harvested energy (in $\mu$J) and is drawn from $\{10,20,30,40,50,60,70,80,90,100\}$.  These values are based on the measurement data reported in \cite{WPT1}. As for data packets, each state corresponds to the data rates of IEEE 802.11g.  Each state has a uniform probability to enter another state.
At each time $t$, packet arrivals follow a Bernoulli process.  When approximating the value of a given state, i.e., $\bar{V}_t()$, we sample $N=100$ states. We set $T=10000$ and calculate the average queue length and harvested energy over 20 runs. We set $\gamma=1$, recorded the average queue lengths at APs and also energy harvested by all sensor nodes.

We derive matrix $\mathbf{A}$ as follows.  Each AP has a unique identity (ID).  For a given transmission set $\tau$, we start with an AP $a$ with the lowest ID.  We then greedily add another AP into the set $\tau$ if it does not interfere with APs in $\tau$.  This continues until no more APs can be added.  We then add $\tau$ into $\mathbf{A}$ if it is new.  Otherwise, we construct a new $\tau$ and repeat the process by starting from the AP with the next highest ID.
Note, another heuristic can be used to construct the independent sets in matrix $\mathbf{A}$, which may yield a bigger or smaller WLAN capacity region. In other words, the new heuristic only scales our results and does not alter our conclusions.

\begin{figure}[ht]
\centering
\subfigure[]{
    \includegraphics[width=6.2cm]{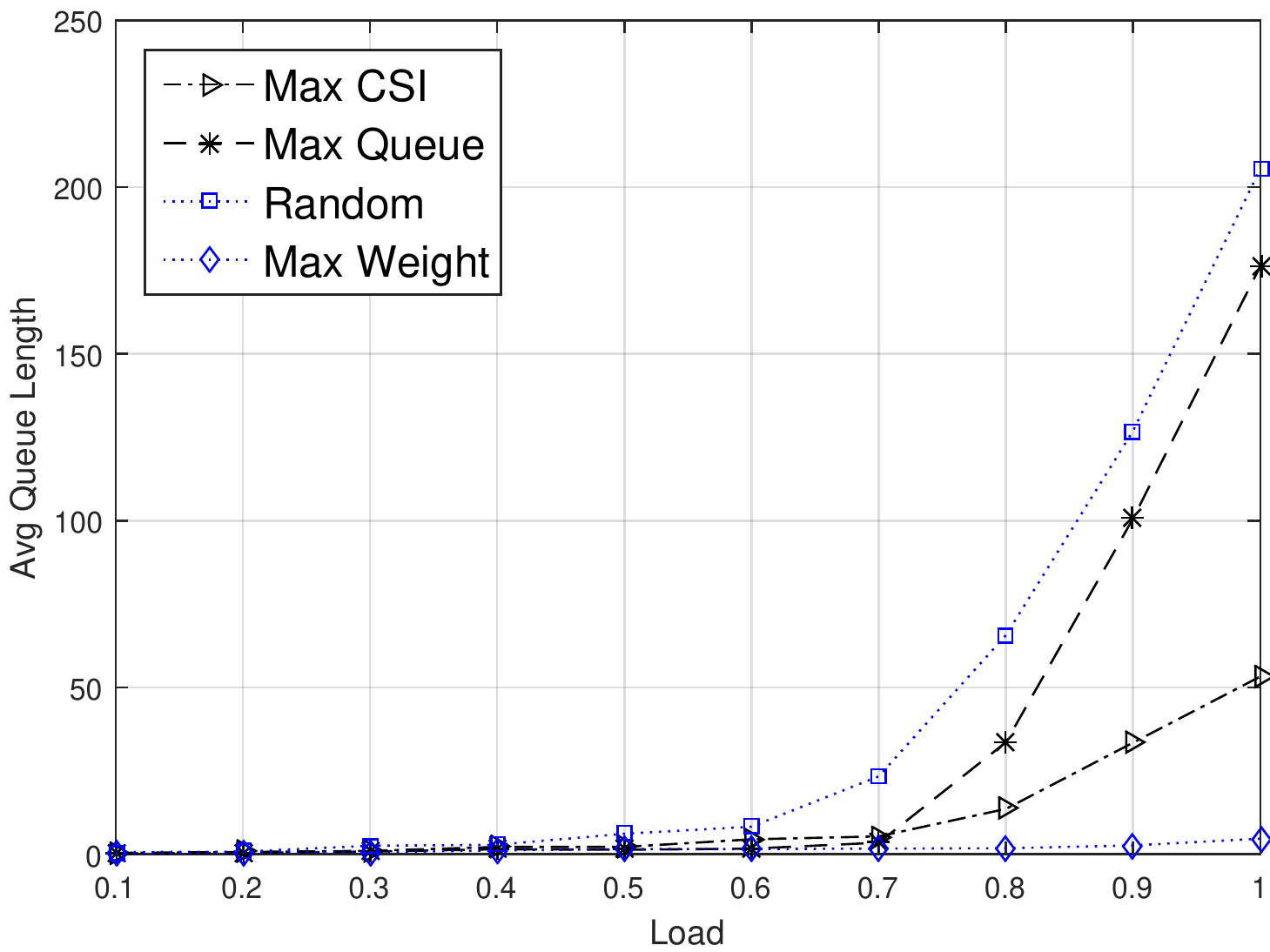}
    \label{fig:subfig1}
}\hspace{-0.5em}
\subfigure[]{
    \includegraphics[width=6.2cm]{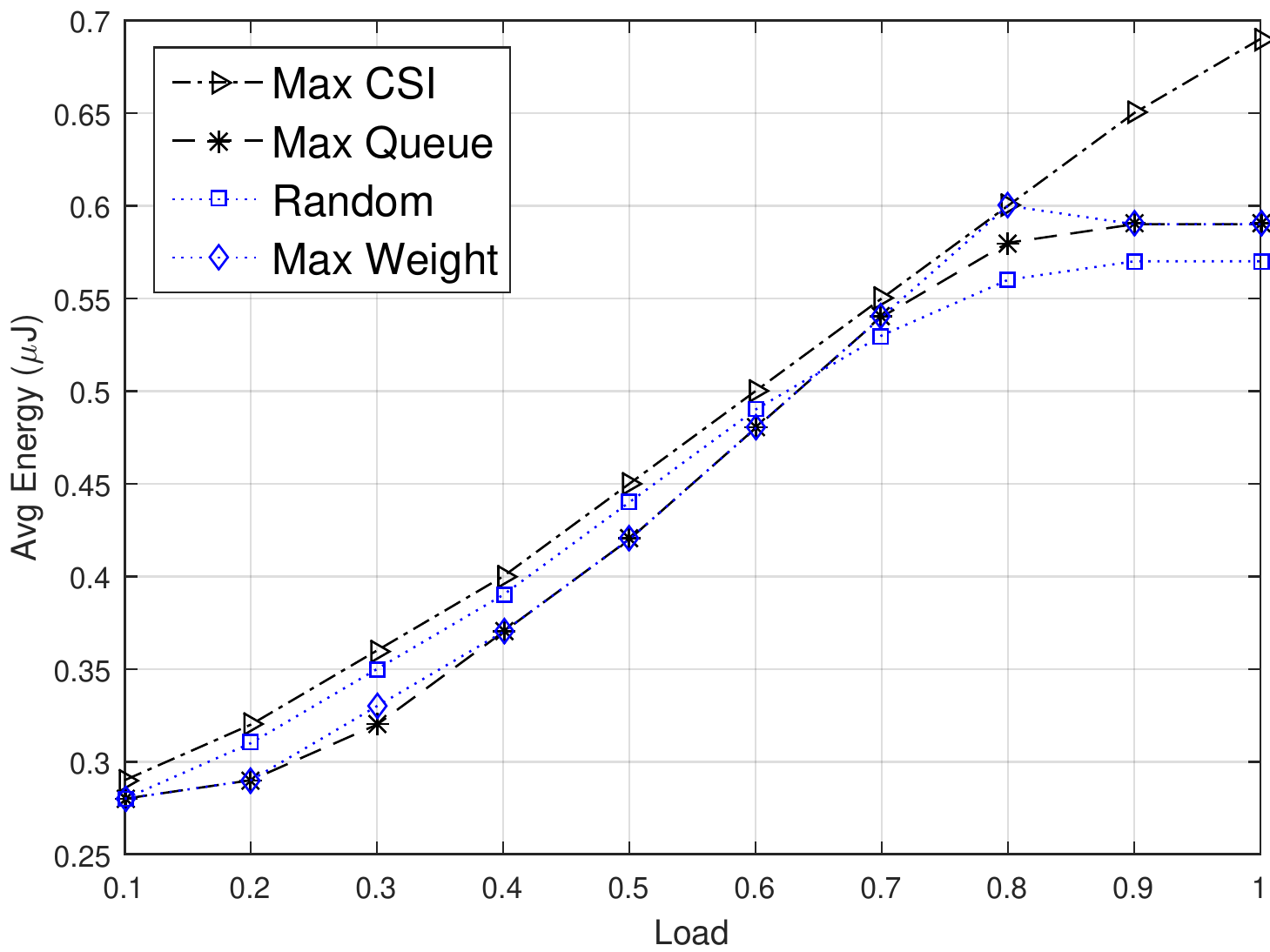}
    \label{fig:subfig2}
}
\caption[]{Probability of arrivals versus average (a) queue length and (b) harvested energy}
\label{FIG2ab}
\end{figure}

Figure \ref{FIG2ab}(a) shows that the max weight policy is able to support the highest arrival rates.  This is because APs pick the longest queue that can fully take advantage of the channel condition at each time $t$.  In contrast, policies such as random and max queue may choose a station with few packets or poor channel.
In terms of harvested energy, see Figure \ref{FIG2ab}(b), when the load is high, max CSI allows APs to transmit the highest number of packets, and thereby, allow sensor nodes to harvest more energy.
At lower loads, the max weight policy may cause APs to choose a station with few packets despite having good channel conditions.  Hence, the actual number of transmitted packets is low.
Lastly, we observe that larger transmission sets, those with more transmitting APs, are preferred because they yield higher total reward.  Hence, some APs may receive fewer transmission opportunities in the short term.
%

\begin{figure}[ht]
\centering
\subfigure[]{
    \includegraphics[width=6.2cm]{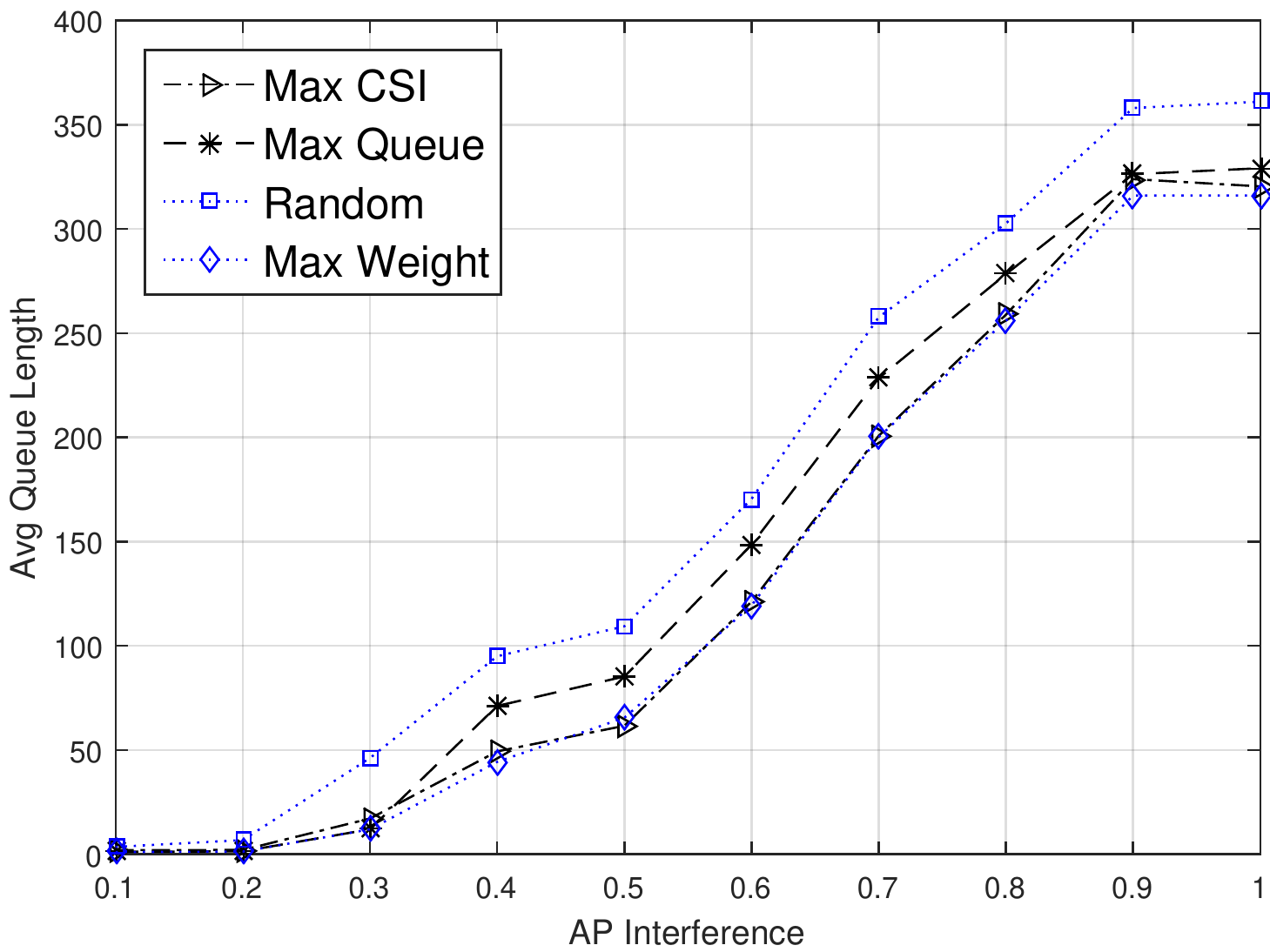}
    \label{fig2:subfig1}
}\hspace{-0.5em}
\subfigure[]{
    \includegraphics[width=6.2cm]{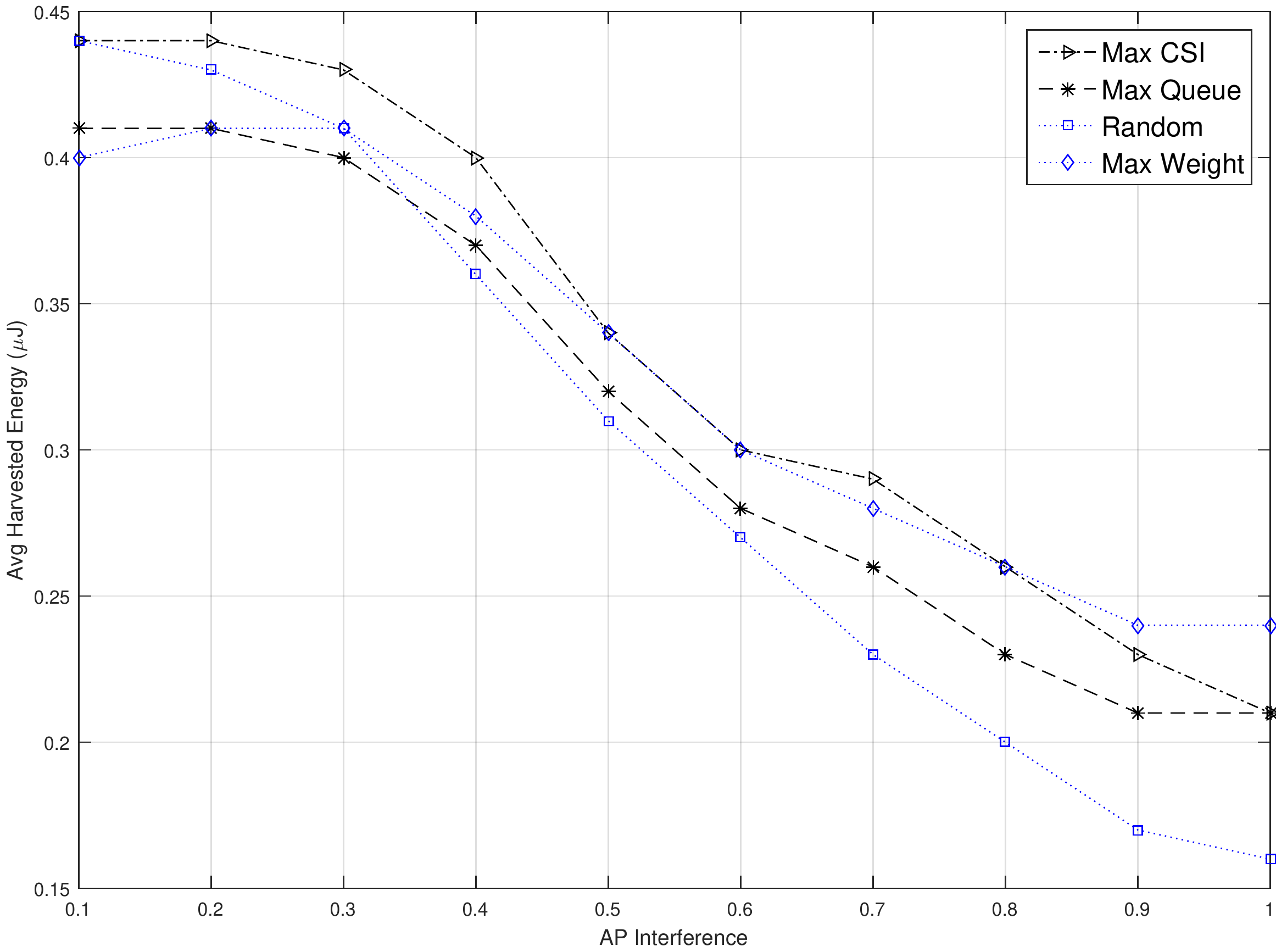}
    \label{fig2:subfig2}
}
\caption[]{The impact of AP interference versus the average (a) queue lengths, and (b) harvested energy}
\label{FIG3ab}
\end{figure}

Next, we consider increasing inter-APs interference.   The traffic load is fixed at 0.5.
When the interference is low, more APs are able to transmit simultaneously.  On the other hand, when interference is high, only one AP out of $|\mathcal{A}|$ may transmit at a time.
Figure \ref{FIG3ab}(a) shows that the max weight policy has the best performance due to the previously discussed reasons.
The amount of harvested energy decreases when more APs interfere with one another; see Figure \ref{FIG3ab}(b).  This is because each AP transmits infrequently.
At a traffic load of 0.5, APs usually have packets awaiting transmission.  Thus, both the max weight and CSI policies are able to fully exploit good channel conditions.  In contrast, the random and max queue policies may select a station with poor channel conditions.


\begin{figure}[ht]
\centering
\subfigure[]{
    \includegraphics[width=6.2cm]{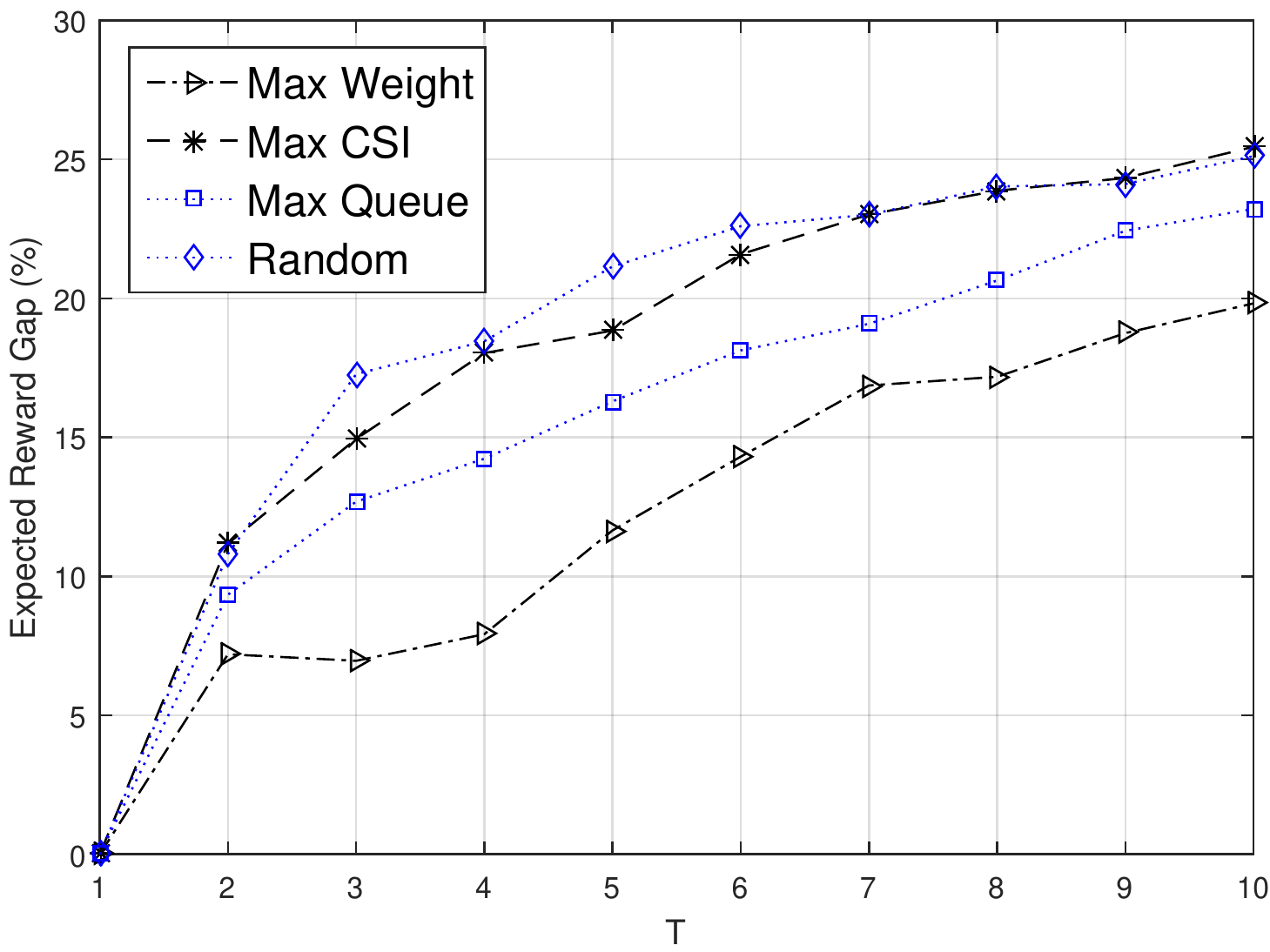}
    \label{fig4ab:subfig1}
}\hspace{-0.5em}
\subfigure[]{
    \includegraphics[width=6.2cm]{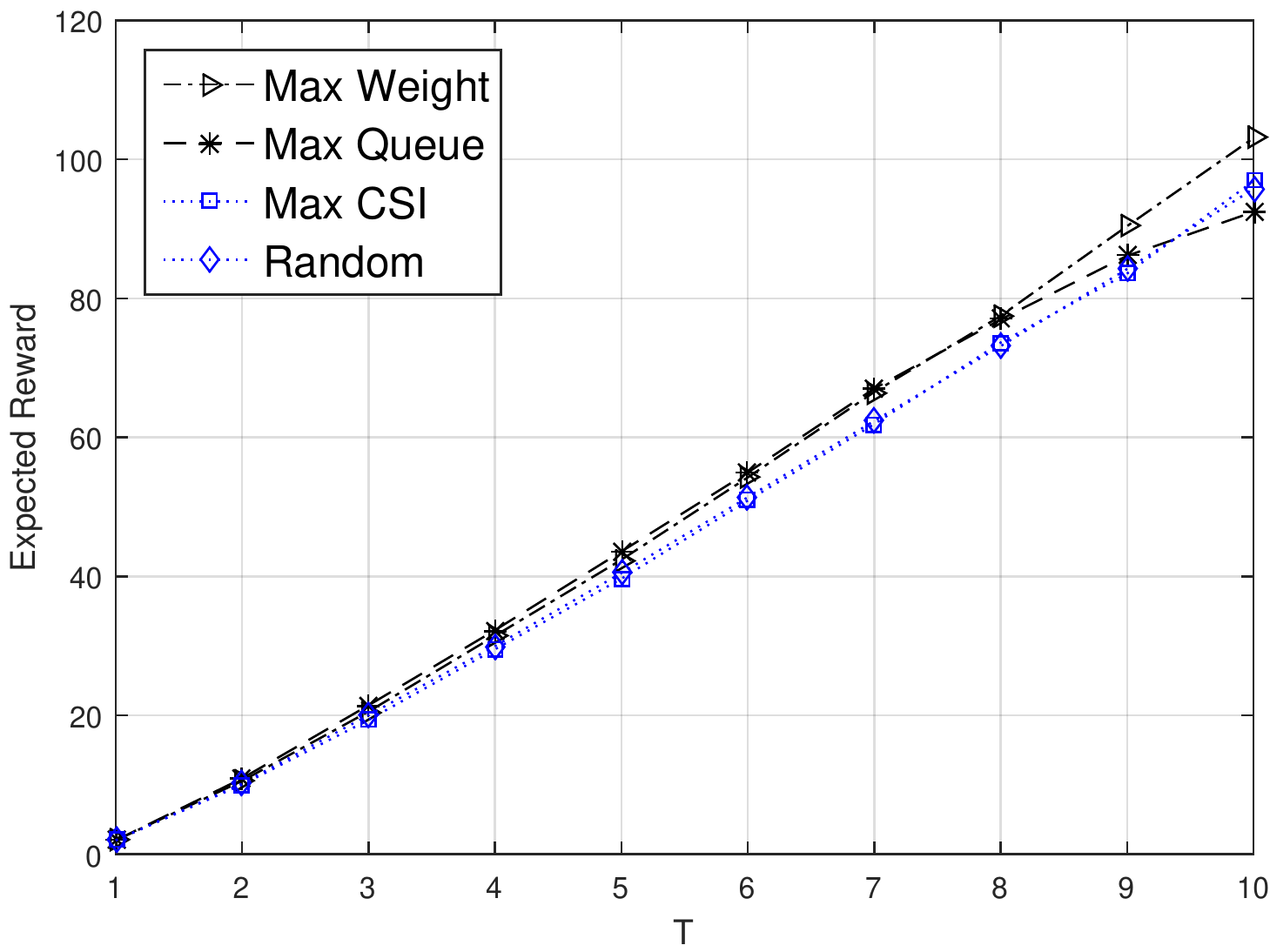}
    \label{fig4ab:subfig2}
}
\caption[]{(a) Average gap between approximated and exact value, and (b) exact reward attained by tested policies}
\label{FIG4ab}
\end{figure}

Lastly, we study the case where Equ. (\ref{MDPVAL}) can be solved exactly as opposed to being approximated; see Equ. (\ref{MDPVALAPPROX}).  Note that finding a good method that has minimal or no gap between the approximate and exact value remains an open research question; see \cite{PowellApproxDP}.  In our case, we also have the added complexity of deriving the set of actions or transmission sets, which involves finding the maximum independent set; an NP-hard optimization problem.
To this end, we consider a small scenario that allows us to generate {\em all} states and also the optimal transmission sets.  Specifically, the scenario in question has two interfering APs.  Hence, there are only two transmission sets, each with one transmitting AP. Each AP has two stations and one sensor node.  The queue length takes on three states: \{0,1,2\}.  There are two channel states, namely \{1,2\} and in each state, there are either zero, one or two packet arrivals.  In total we have 5184 states.  When computing Eq.  (\ref{MDPVALAPPROX}), we set $N=1000$.  
Figure \ref{FIG4ab}(a) shows the gap or percentage from the exact expected reward value.
We see that with increasing $T$, the gap increases and it is advantageous to keep $T$ small; this ensures a smaller gap and also fast computation time. Advantageously, we see that the max weight policy has the smallest gap.
Note that for a given $T$, the gap is a constant.  Note, we have tried higher values of $N$, at the expense of longer computation time, but the results remain similar. 
Figure \ref{FIG4ab}(b) shows the exact reward value obtained by the tested policies.  The trend is consistent with earlier results whereby the max weight policy has the best expected reward.  

\section{Conclusion}
\label{CONC}
Future dense deployment of APs are likely to be used to charge nearby energy-harvesting devices and also deliver data to stations simultaneously.  
To this end, we use a MDP to study various policies and show that the maximum weight policy ensures APs have short queues whilst ensuring energy-harvesting devices receive ample energy.  Moreover, it yields the smallest gap between the approximate and exact value.
%

\bibliographystyle{IEEEtran}
\bibliography{WPT}

\end{document}